\documentclass[twocolumn, showpacs, amsmath, amssymb, prd]{revtex4}
\usepackage{graphicx}
\usepackage{dcolumn}
\usepackage{bm}

\begin{document}

\title{ Vector potentials in gauge theories in flat spacetime } 
\author{C. W. Wong}
\email{cwong@physics.ucla.edu}

\affiliation{Department of Physics and Astronomy, University of California, 
Los Angeles, CA 90095-1547, USA}

\begin{abstract}
A recent suggestion that vector potentials in electrodynamics (ED) are nontensorial objects under 4D frame
rotations is found to be both unnecessary and confusing. As traditionally used in ED, a vector potential 
$A$ always transforms homogeneously under 4D rotations in spacetime, but if the gauge is changed by the 
rotation, one can restore the gauge back to the original gauge by adding an inhomogeneous term. It is 
then ``not a 4-vector'', but two: one for rotation and one for translation. For such a gauge, it
is much more important to preserve {\it explicit} homogeneous Lorentz covariance by simply skipping the 
troublesome gauge-restoration step. A gauge-independent separation of $A$ into a dynamical 
term and a non-dynamical term in Abelian gauge theories is re-defined more generally as the terms caused 
by the presence and absence respectively of the 4-current term in the inhomogeneous Maxwell equations for 
$A$. Such a separation {\it cannot} in general be extended to non-Abelian theories where $A$ satisfies 
nonlinear differential equations. However, in the linearized iterative solution that is perturbation 
theory, the usual Abelian quantizations in the usual gauges can be used. Some nonlinear complications 
are briefly reviewed. 
\end{abstract}

\maketitle

\section{Introduction}

Lorc\'e \cite{Lorce13} has recently given, among many interesting results, a nonstandard description of 
the vector potentials $A_{\rm L}$ in Lorenz (L) gauges in classical electrodynamics (CED). His motivation 
is to be able to claim that the CED vector potential $A(x)$ in {\it any} gauge is Lorentz covariant 
under {\it four-dimensional} (4D) frame rotations. He does this by re-defining homogeneous Lorentz 
covariance for $A(x)$, but not for $x$ itself, as {\it inhomogeneous} Lorentz covariance where an 
inhomogeneous term for 4D translation can be added to the result of a 4D rotation. He further claims that 
the Lorenz gauge is ``nothing special'' because ``one cannot conclude that the only possible Lorentz 
transformation law for the gauge potential is the four-vector one, unless one removes the residual gauge 
freedom'', and that vector potentials are ``nontensorial objects'' \cite{Lorce13}. 

One objective of this paper is to explain in Sect. \ref{sec: hLT} why Lorc\'e's revisionist view is both 
unnecessary and confusing 
compared to the traditional textbook language (\cite{Bjorken65} -- \cite{Panofsky55}, for example) that is 
already simple, clear and consistent. In this standard language using special relativity in flat space (and 
not general relativity in curved space), 4-vectors are {\it defined} to transform homogeneously under 4D 
frame rotations, just like $x$ itself. For $A(x)$ in gauge theories, the central issue turns out to be 
whether its chosen gauge in frame $x$ will remain unchanged after a 4D rotation, i.e., a {\it homogeneous 
Lorentz} (hL) transformation. We shall show that the hL transformation of $A(x)$ is gauge-preserving 
in {\it covariant gauges} (cg), and gauge non-preserving in {\it non-covariant} (nc) gauges. 

If the gauge is changed for a vector potential $A(x)$ in the rotated frame $x'$, its altered gauge can be 
restored back to the original gauge used in frame $x$ by adding a change-of-gauge term to give a total 
gauge-restored but inhomogeneous Lorentz transformation. Such an $A(x)$ is described correctly as ``not a 
4-vector'' in textbooks \cite{Bjorken65, Weinberg96} because it is the sum of {\it two} 4-vectors, one 
for rotation and one for translation. However, one can easily preserve the important requirement of 
{\it explicit} homogeneous Lorentz covariance by simply skipping this troublesome gauge-restoration step. 
Some implications of this change of perspective are discussed. 

The second objective of this paper is to show in Sect. \ref{sec: Dyn/Ndy} that the vector potential $A$ in 
ED is non-unique in general by using the separation $A = A_{\rm dyn} + A_{\rm ndy}$ known from the theory 
of linear differential equations. Here the {\it dynamical} (dyn) part $A_{\rm dyn}$ and {\it non-dynamical} 
(ndy) part $A_{\rm ndy}$ are respectively the solutions of the Maxwell equations for $A$ with and without 
the 4-current density $j$. This separation further highlights the important role played by the 
{\it Coulomb} (C) gauge in clarifying a different known structure of $A$ present in both dynamical and 
non-dynamical parts: The Coulomb gauge transversality condition $\bm{\nabla} \cdot {\bf A}_{\rm C} = 0$ 
selects a transverse ($\scriptstyle\perp$) or physical (phys) part ${\bf A}_{\rm C} = {\bf A}_\perp = 
{\bf A}_{\rm phys}$ that is known to be gauge-invariant (\cite{CT89}, for example). It excludes the 
longitudinal ($\scriptstyle\parallel$) or ``pure-gauge'' part ${\bf A}_\parallel = {\bf A}_{\rm pure}$ due 
to gauge transformations known to contribute nothing to the field tensor. The sum 
$A = A_{\rm phys} + A_{\rm pure}$ is thus {\it explicitly} gauge invariant, as shown by Chen {\it et al.} 
\cite{Chen08}. 

The dynamical/non-dynamical treatment is naturally gauge invariant in ED. The vector potential $A$ 
is non-unique because a non-dynamical part $A_{\rm ndy}$ can be added to any particular solution to 
generate other solutions. $A_{\rm ndy}$ also has a transverse/longitudinal (or physical/gauge) 
decomposition: Its gauge part is the entire gauge part of $A$, and has no physical consequences. Its 
physical part generates nonzero field tensors $F^{\mu\nu}$ describing transverse electromagnetic waves 
in free space and longitudinal electric fields from scalar potentials satisfying the Laplace equation 
for different boundary conditions. However, in applications such as \cite{Chen08}, where one is only 
interested in one particular solution $A_{\rm dyn}$ for the gauge field bound inside a distinct atomic 
state by a unique bound-state boundary condition, there is no need to add a further $A_{\rm ndy}$ term 
of the physical type. The dynamical/non-dynamical treatment can accommodate these special cases too.   

The decomposition of $A$ in ED into parts cannot in general be extended to non-Abelian theories where 
$A$ satisfies nonlinear differential equations. Even the 4-current and field tensor are themselves 
gauge dependent. However, a linearized iterative treatment using perturbation theory permits Abelian 
quantizations in the usual gauges as an approximation. Some nonlinear complications of non-Abelian gauge 
theories are reviewed, especially from the perspective that second quantization is a linearization 
process that {\it unavoidably} leaves most of the native nonlinearity untreated until after quantization.

\section{Homogeneous Lorentz covariance}
\label{sec: hLT}

Recall that a (single-valued) vector field $A(x)$ in an inertial frame $x$ of a flat and isotropic 
spacetime is one where its value at point $x$ is a 4-vector, i.e., a 4D ``oriented arrow'' 
of definite length (figuratively speaking) present {\it in spacetime} itself. A 4-vector as used here is
{\it any} 4-component object whose components are defined by the same coordinate axes as the frame $x$ 
and are compatible in physical unit so that its ``length'' can be calculated. Under a homogeneous 
Lorentz (hL) transformation (i.e., a 4D rotation) of the coordinate frame defined by $x'=\Lambda x$, 
such a 4D vector field transforms as 
\begin{eqnarray}
A'(x'=\Lambda x) &=& \Lambda A(x), \quad {\rm because} \nonumber \\
A &=& e'_\mu(x') A'^\mu(x') = e_\nu(x) A^\nu(x): \nonumber \\
A'^\mu &=&  e'^\mu \cdot A = (e'^\mu \cdot e_\nu) A^\nu \nonumber \\
&=& \Lambda^\mu_\nu A^\nu,
\label{LTvf} 
\end{eqnarray}
where $e_\nu(x)$ is a unit vector. This hL transformation gives the new components $A'^\mu$ in the rotated 
frame $x'$ of each spacetime arrow $A(x)$ of unchanged length and orientation. The rotation matrix 
$\Lambda$ on the right-hand side (RHS) for position $x \neq 0$ uses a local $x$ frame (a copy of the 
$x$-frame with its origin translated to position $x$), and not the original $x$-frame centered at $x=0$. 
In other words, a real 4-vector field is {\it defined} to be based on the 4D real-number system that 
transforms as $x' = \Lambda x$ under 4D frame rotations. It can be generalized to a complex 4-vector field by 
adding an overall phase factor at each position $x$. 

Each 4D vector in $A(x)$ defined by Eq. (\ref{LTvf}) will be called an hL {\it covariant 4-vector} in 
this paper, meaning a 4-vector under 4D frame rotations. This is the same object as Lorc\'e's ``Lorentz 
4-vector'' or the usual ``4-vector'' of textbooks \cite{Bjorken65}--\cite{Panofsky55}. Tensors built from 
covariant 4-vectors are covariant 4-tensors, while an invariant scalar has the same single value in all hL frames. The 
language used here and in textbooks is thus the simplest generalization of 3D spatial vector fields in 
Euclidean space of signature (3,0) to 4D spacetime vector fields in Euclidean space of signature (3,1), 
where the space-like specification comes first, irrespective of the overall sign convention used.

In Lorc\'e's revised language \cite{Lorce13, Leader14, Lorce15}, 4D rotational covariance of $A(x)$ is 
re-defined as an {\it inhomogeneous} Lorentz transformation, containing terms for both rotation between 
two frames and translation in a single frame, even though the most important 4-vector, namely $x$ itself, 
satisfies only an hL transformation. (This exception is needed because the associated inhomogeneous Lorentz 
transformation for $x$ itself defines a full-blown Poincar\'e transformation that is not the subject 
under discussion.) This inconsistent treatment is the most serious source of confusion in the proposed 
revision.

A prime example of the standard usage adopted in this paper appears in the definition of the gauge covariant 
derivative in QED involving particles of charge $e$ ($e = -|e|$ for electrons):
\begin{eqnarray}
D^\mu \equiv \partial^\mu + ieA^\mu.
\label{D} 
\end{eqnarray}
In gauge theories of interaction, it is the requirement of local gauge invariance of the Lagrangian under 
the local gauge transformation (LGT) of the complex Dirac field 
$\psi(x)  \longrightarrow^{\rm LGT} e^{-ie\omega(x)}\psi(x)$ for a charged particle that forces the 
introduction of the gauge potential $A(x) = \partial \omega(x)$, where $\omega(x)$ is a real scalar field. 
Thus $A$ by design is the same covariant 4-vector object as $\partial$ in flat space. The fact that this 
$A$ appears in the invariant scalar product $\overline{\psi}\gamma_\mu D^\mu \psi$ in QED is what gives 
the interaction Lagrangian $\overline{\psi}\gamma_\mu A^\mu\psi = j\cdot A$ (where 
$j = \overline{\psi}\gamma\psi$) both its interaction and its explicit hL invariance. This $A^\mu$ is 
thus intended to be a covariant 4-vector. We shall explain why it sometimes turns out to be ``not a 
4-vector'' thus ruining the explicit hL invariance of the interaction Lagrangian, and how it can be chosen 
to be a covariant 4-vector for {\it any} choice of gauge in frame $x$ for gauge-invariant theories in flat 
spacetime. 

Recall that in CED alone, the electromagnetic field tensor 
$F^{\mu\nu} = \partial^\mu A^\nu - \partial^\nu A^\mu$ in frame $x$ is unchanged by the {\it gauge 
transformation} (GT) :
\begin{eqnarray}
A^\mu(x) \longrightarrow^{\rm GT} \, A^\mu(x) + \,\partial^\mu \Omega(x),
\label{GT} 
\end{eqnarray}
where $\Omega(x)$ is a real scalar field. The vector potential $A$ introduced by Eq. (\ref{D}) is thus 
highly non-unique. Part of the resulting redundant gauge degrees of freedom can be eliminated 
by imposing a constraint called a gauge condition, expressible in the form 
\begin{eqnarray}
{\cal G}\{A(x)\} \equiv C(x)\cdot A(x) = 0,
\label{Gcond} 
\end{eqnarray}
where $C(x)$ is a chosen covariant 4-vector operator or field. Under a general hL transformation, the chosen 
gauge can be either (a) covariant or gauge preserving, and denoted {\it covariant gauge} (cg), or 
(b) non-covariant or gauge non-preserving, denoted {\it variable gauge} (vg):
\begin{eqnarray}
{\rm (a) \quad If \;}\Lambda C(x) = C'(x'): && \nonumber \\
 C(x)\cdot A_{\rm cg}(x) &=& C(x)\cdot \Lambda^{-1}\Lambda A_{\rm cg}(x), \nonumber \\
  &=& C'(x')\cdot A'_{\rm cg}(x'), \nonumber \\
  A'_{\rm cg}(x') &=& \Lambda  A_{\rm cg}(x). \nonumber \\
{\rm (b) \quad If \;} \Lambda C(x) \neq C'(x'): &&  \nonumber \\
C(x)\cdot A_{\rm vg}(x) &\neq & C'(x')\cdot [\Lambda A_{\rm vg}(x)]. \quad
\label{cg/vg}
\end{eqnarray}
(The identity $C\cdot \Lambda^{-1} = [\Lambda C]\cdot$ has been used in Eq. (\ref{cg/vg}).) Thus in the 
standard language, the gauge-preserving covariant 4-vector property of $A_{\rm cg}(x)$ in any covariant 
gauge cg described in Eq. (\ref{cg/vg}a) is a consequence of the hL invariance of its gauge condition.  

The inhomogeneous Maxwell equation for $A$ in ED is
\begin{eqnarray}
\partial_\mu(\partial^\mu A^\nu - \partial^\nu A^\mu) 
&=&  \partial^2 A^\nu - \partial^\nu(\partial \cdot A) \nonumber \\
\equiv {\cal L}_\mu^\nu A^\mu &=& j^\nu, 
\label{inhMaxEq}
\end{eqnarray}
where ${\cal L} = {\cal L}(x)$ is a {\it gauge-dependent} linear differential operator and $j = j(x)$ is a 
gauge-independent covariant 4-vector in spacetime (because it is made up of point charges moving with 
4-velocities in frame $x$ \cite{Jackson99}). For a covariant gauge cg where $A_{\rm cg}(x)$ is a 
covariant 4-vector, the full covariant tensor structure of the rank-2 covariant 4-tensor ${\cal L}$ is 
preserved under 4D rotations. So covariant gauges are indeed special. 

Among these cg gauges, the Lorenz gauge is the most special and indeed unique because 
its gauge condition $\partial \cdot A_{\rm L}(x) = 0$ allows Eq. (\ref{inhMaxEq}) to be simplified to 
\begin{eqnarray}
\partial^2 A_{\rm L} = j.
\label{inhWaveEq}
\end{eqnarray}
Since ${\cal L}_{\rm L} = \partial^2$ is now an invariant scalar operator, the covariant 4-vector nature of 
$A_{\rm L}$ is dictated by the covariant 4-vector nature of the 4-current $j$ on the RHS \cite{Panofsky55}, 
even for the special case $j=0 = (0,0,0,0)$. So the covariant 4-vector property of {\it every} solution 
$A_{\rm L}$ for both $j \neq 0$ and $j = 0$ is a consequence of the Lorenz condition alone. 

Further elaboration of the covariant 4-vector nature of $A_{\rm L}$ might be helpful: First define 
$A_{\rm L}(x)$ as the multivalued set, object or ``function'' containing all the multiple solutions of 
Eq. (\ref{inhWaveEq}) as its multiple values. Then display its behavior under 4D rotation of the external 
local frame $x$ to $x'$:
\begin{eqnarray}
\Lambda \partial^2 A_{\rm L}(x) &=&  (\Lambda \partial^2 \Lambda^{-1}) \,(\Lambda A_{\rm L}(x))
\nonumber \\
= \partial'^2 A'_{\rm L}(x') &=&\Lambda j(x) = j'(x'),
\label{hLTwaveEq}
\end{eqnarray}
where $A'_{\rm L}(x') = \Lambda A_{\rm L}(x)$ from Eq. (\ref{cg/vg}a) has been used. So the wave equations
(\ref{inhWaveEq}) transform covariantly for every solution contained in $A_{\rm L}(x)$ for the simple reason 
that the multivalued object $A'_{\rm L}(x')$ is exactly the same object as the original $A_{\rm L}(x)$, 
but with components decomposed relative to the rotated local frame $x'$. This covariance refers initially
to the constancy of the 4-component object $A_{\rm L}(x)$ under {\it external} frame rotations. 

However, ED as an hL invariant theory in isotropic spacetime contains an additional symmetry of 
importance in physics: Spacetime itself shows no 4D orientation preference for ED phenomena, so 
that only the relative orientation between frame and solution is physically 
meaningful. There thus exists a solution $A_{\rm L}(x')$ in frame $x$ where the solution has been rotated 
in the opposite direction and is numerically indistinguishable from $A'_{\rm L}(x')$. The covariance of 
Eq. (\ref{hLTwaveEq}) can then be {\it interpreted} as referring to such rotated solutions in hL invariant
theories. In theories that are not even implicitly hL invariant, such an interpretation is not admissible.

In Eq. (\ref{cg/vg}b) on the other hand, $\Lambda A_{\rm vg}(x)$ satisfies the gauge $\Lambda C(x)$ that 
differs from $C'(x')$; it is thus a gauge non-preserving covariant 4-vector. {\it If} one insists on 
restoring the gauge in $x'$ from gauge $\Lambda C$ back to the original non-covariant (nc) gauge, it will 
be necessary in Abelian theories to {\it add} an extra inhomogeneous 4-vector term 
$R'_{\Lambda C \rightarrow {\rm nc}}(x') \equiv \partial' \Omega_{\Lambda C \rightarrow {\rm nc}}(x') $ 
for gauge restoration to give the inhomogeneous transformation 
\begin{eqnarray}
A_{\rm nc}'(x'=\Lambda x) &=& \Lambda A_{\rm nc}(x) +  R'_{\Lambda C \rightarrow {\rm nc}}(x'), 	
\nonumber \\ {\rm where} \qquad {\cal G}_{\rm nc}\{ A'_{\rm nc}(x') \} &=& 0
\label{restore} 
\end{eqnarray}
is the restored gauge condition, and $A_{\rm nc}(x) = A_{\rm vg}(x)$ has been re-named ``nc'' for greater 
clarity. Note that the extra term $R'_{\Lambda C \rightarrow {\rm nc}}$ is {\it not} concerned with the 
{\it residual} gauge degree of freedom describing the non-uniqueness of $A'_{\rm g}$ itself in a single 
gauge g in frame $x'$ for the same field tensor $F$. 

The new term $R'_{\Lambda C \rightarrow {\rm nc}}$ in Eq. (\ref{restore}) ruins the hL covariance property 
of $\Lambda A_{\rm nc}(x)$ however, because the four spacetime components are treated asymmetrically in 
non-covariant gauges. An asymmetry then appears in the differential operator ${\cal L}$ on the LHS of Eq. 
(\ref{inhMaxEq}) for a non-covariant gauge. So after gauge restoration, $A_{\rm nc}'(x')$ of Eq. 
(\ref{restore}) is no longer a covariant 4-vector, but an hL {\it non-covariant} 4-vector. This is the 
standard picture described in textbooks \cite{Bjorken65, Weinberg96, CT89}. See also \cite{Waka12}. In 
particular, the interaction Lagrangian 
$j\cdot A_{\rm nc} = (\Lambda j) \cdot (\Lambda A_{\rm nc}) \neq  j' \cdot  A'_{\rm nc}$, where the 
4-current $j$ remains a covariant 4-vector, is no longer explicitly hL invariant. The DE (\ref{inhMaxEq}) too 
is no longer covariant even though all tensor indices correctly describe matrix multiplications, because 
$\Lambda {\cal L}_{\rm nc}(x) \Lambda^{-1} \neq {\cal L}'_{\rm nc}(x')$: The rotated solutions in frame 
$x$ also do not solve the same Eq. (\ref{inhMaxEq}) with ${\cal L}_{\rm nc}$.  

If a gauge-restored but non-covariant 4-vector potential $A_{\rm nc}$ is used in a {\it gauge-invariant} 
formulation of ED, it must contain hidden hL covariance because one can always gauge transform to the 
Lorenz gauge (or any other covariant gauge cg) where the hL covariance of $A_{\rm cg}$ can be explicitly 
displayed. 

How can this hidden hL covariance be made explicit? This question can be answered by using the sequential 
(gauge $\rightarrow$ Lorentz $\rightarrow$ gauge) transformations \cite{Wong10}: 
\, $ A_{\rm nc}(x) \longrightarrow^{\rm GT} \, A_{\rm cg}(x) \longrightarrow^{\rm LT} \, A'_{\rm cg}(x')
\, \longrightarrow^{\rm GT} \,  A'_{\rm nc}(x') $, where $x$ defines any initial inertial frame and 
$x' = \Lambda x$ is the new hL (4D rotated) frame. The last three $A$s have the interesting structure 
\begin{eqnarray}
{\rm (a)} \qquad \qquad A_{\rm cg}(x) &=& A_{\rm nc}(x) + R_{{\rm nc} \rightarrow {\rm cg}}(x), 
\quad {\rm where} \quad \nonumber \\
R_{{\rm nc} \rightarrow {\rm cg}}(x) &=& \partial \Omega_{{\rm nc} \rightarrow {\rm cg}}(x); \nonumber \\
{\rm (b)} \quad \hspace{0.05pt} A'_{\rm cg}(x'=\Lambda x) &=& \Lambda A_{\rm cg}(x) \nonumber \\
               &=& \Lambda \left[ A_{\rm nc}(x) + R_{{\rm nc} \rightarrow {\rm cg}}(x) \right]; \nonumber \\
{\rm (c)} \qquad \quad \;\; \hspace{0.25pt} 
A'_{\rm nc}(x') &=& A'_{\rm cg}(x') + R'_{{\rm cg} \rightarrow {\rm nc}}(x'). 
\label{sequential}
\end{eqnarray}
Here $\Lambda \equiv \Lambda(x, x')$. These expressions can be combined to give the gauge-restored Lorentz 
transformation relation for the non-covariant but fixed-gauge 4-vector $A_{\rm nc}$
\begin{eqnarray}
A'_{\rm nc}(x') &=& \Lambda A_{\rm nc}(x) + R'_{\Lambda C \rightarrow {\rm nc}} (x, x'), 
\label{combined} \\
R'_{\Lambda C  \rightarrow {\rm nc}} (x, x') &=& R'_{{\rm cg} \rightarrow {\rm nc}}(x') 
+ \Lambda R_{{\rm nc} \rightarrow {\rm cg}}(x). \quad
\label{error}
\end{eqnarray}
Eq. ({\ref{combined}) is a refinement of Eq. (\ref{restore}) for the same end result, namely an 
inhomogeneous transformation in the 4D functional space of $A$ involving two hL-related local frames $x$ 
and $x'$ in 4D spacetime with no translation between them. 

All the gauge transformations involved in Eqs. (\ref{restore}, \ref{sequential}) are change-of-gauge ones; 
none is concerned only with a residual gauge term causing no gauge change. However, a necessarily 
longitudinal residual gauge term $R_{\rm g}(x) \equiv \partial\Omega_{\rm g}(x)$ or $R'_{\rm g}(x')$ can 
in general be added to a vector potential in any {\it non-transverse} gauge g and in the frame $x$ or $x'$ 
in these equations without changing their validity. This $R_{\rm g}$ is by definition gauge-preserving. 
It is often called a gauge transformation of the second kind \cite{Sakurai67}. In Lorenz gauges for example, 
the term satisfies the wave equation $\partial^2 \Omega_{\rm L} = 0$, or $\partial^2 R_{\rm L} = 0$.

Since ED is a linear theory in $A$, an ED gauge condition ${\cal G}_{\rm g}\{A\} = 0$ is almost always one 
satisfying the linearity property
\begin{eqnarray}
{\cal G}_{\rm g}\{ A_{\rm g}(x) + R_{\rm g}(x) \} 
&=&  {\cal G}_{\rm g}\{A_{\rm g}(x) \} + {\cal G}_{\rm g}\{ R_{\rm g}(x) \} \nonumber \\
&=& 0,
\label{linear}
\end{eqnarray}
where $R_{\rm g}(x)$ is a residual gauge term. This linearity has the consequence that 
$A_{\rm g}$ and $R_{\rm g}$ separately or together satisfies the gauge condition. Since the gauge 
condition also dictates the hL covariance property of vector potentials, $A_{\rm g}$ and $R_{\rm g}$ 
separately or together must be only covariant 4-vectors, or only non-covariant 4-vectors. This means 
that all residual gauge terms can simply be {\it absorbed} into their parent terms (such as $A_{\rm g}$ 
in Eq. (\ref{linear})) and not shown explicitly, if each $A_{\rm g}$ denotes a multivalued object 
containing {\it all possible values} allowed by the residual gauge degree of freedom. 

For the covariant Lorenz gauge for example, the hL transform $A'_{\rm L}(x') = \Lambda A_{\rm L}(x)$ 
is multivalued if the original $A_{\rm L}(x)$ is multivalued, both containing residual gauge terms. 
On the other hand, for the non-covariant Coulomb gauge where the residual gauge degree of freedom is 
absent, the hL transform $\Lambda  A_{\rm C}(x)$ requires an appropriate multivalued gauge restoration 
$A'_{\rm C}(x') = \Lambda  A_{\rm C}(x) + R'_{\Lambda C \rightarrow {\rm C}}$ to remove all unwanted 
residual gauge terms from $\Lambda  A_{\rm C}(x)$. There is thus also no need for any final gauge 
transformation or gauge rotation of the type discussed in \cite{Lorce13}. The fact that any allowed 
residual gauge degree of freedom has been included in the multivalued object $A_{\rm g}$ will be 
expressed mathematically as Eq. (\ref{nd}) in Sect. \ref{sec: Dyn/Ndy} from a more general perspective.

Why should one remain in the same gauge g in frame $x'$ in a non-covariant gauge? It is good to know 
how to do it, but since the gauge degree of freedom under consideration causes no change in 
$F^{\mu\nu}$ and the classical properties it describes, this gauge degree of freedom can be used to 
enforce not the non-covariant gauge condition, but the hL covariance of the gauge non-preserving 
covariant 4-vector $A_{\rm vg}$. That is, one can simply use the hL transform $\Lambda A_{\rm vg}(x)$ 
of Eq. (\ref{cg/vg}b) alone without adding the troublesome gauge-restoring term 
$R'_{\Lambda C \rightarrow {\rm nc}}(x')$, thus allowing the tensor structure of the inhomogeneous 
Maxwell Eq. (\ref{inhMaxEq}) to retain its usual meaning in flat spacetime. 

For the Coulomb gauge for example, the hL transform $\Lambda {\cal L}_{\rm C}(x) \Lambda^{-1}$ in frame $x'$ in 
the variable-gauge approach differs from the operator ${\cal L}'_{\rm C}(x')$ in the gauge-restored 
approach. In either case, the DE has to be solved with the same operator ${\cal L}_{\rm C}(x)$ in frame $x$ 
only; the solution is then transformed differently to frame $x'$ in different treatments. In the variable-gauge 
treatment, all residual gauge terms $R'_{\rm vg}(x')$ that appear now should also be included. So Eq. 
(\ref{cg/vg}b) can be re-written as the covariant but variable-gauge gauge condition 
$C \cdot A_{\rm vg} = (\Lambda C)\cdot (\Lambda A_{\rm vg}) = 0$ to define a special kind of hL 
invariance/covariance for the original vg=nc gauge in frame $x$. For the Coulomb gauge in any frame $x$, 
the result is a {\it covariant Coulomb} (cC) gauge. It is also a subset of $A_{\rm L}$ of the Lorenz (L) 
gauge whose element for frame $x$ also satisfies the Coulomb gauge condition. This special element can be 
located from any element of $A_{\rm L}(x)$ by the {\it residual} gauge transformation:
\begin{eqnarray}
A^\parallel_{\rm cC}(x) &=& A^\parallel_{\rm L}(x) + \partial^\parallel \Omega_{\rm L}(x)
\nonumber \\
&=& 0.
\label{cCcond}
\end{eqnarray}
The vector potentials for covariant axial and temporal gauges can be similarly located.  

For an hL invariant theory such as ED in flat and isotropic spacetime, physics is independent of the 
choice of frame $x$: All Coulomb-gauge results obtained in any one frame describe the same physics as 
Lorenz gauges. This completes our demonstration that the vector potential $A$ introduced by the 
gauge-covariant derivative (\ref{D}) of Abelian gauge theories can always be chosen to be a covariant 
4-vector in flat space satisfying the hL transformation law $A'(x' = \Lambda x) = \Lambda A(x)$ for any 
gauge in frame $x$ including a non-covriant gauge, thus preserving the {\it explicit} hL of the theory.

To summarize, Lorc\'e's proposed revision of the 4-vector language for flat spacetime is unnecessary 
because the traditional textbook usage in ED is simpler. The proposal is confusing because the parent 
hL transformation of the spacetime 4-coordinate $x$ under 4D rotations of the inertial frame centered at 
$x=0$ is not duplicated by the proposed {\it inhomogeneous} Lorentz transformation of the vector potential 
under 4D rotations of the local frames centered at $x$. Finally, the added inhomogeneous term has nothing 
to do with an hL transformation to another frame, but is instead a change-of-gauge transformation in one 
inertial frame needed to repair or anticipate the gauge damage caused by the hL transformation. 

The last comment applies even to non-Abelian gauge theories in flat spacetime where $A(x) = A_a(x)t_a$ is a 
sum over {\it internal} components $A_a(x)$ associated with the $a$-th generators $t_a$ of the gauge group. 
What has changed in non-Abelian (nAb) theories is {\it not} the 4D rotation of the external frame, {\it but} 
their gauge transformation (GT) at point $x$ in frame $x$ \cite{Peskin95}: 
\begin{eqnarray}
A_a^\mu \longrightarrow^{\rm nAbGT} \, A_a^\mu + \,\partial^\mu \Omega_a
+ g f_{abc}A_b^\mu \Omega_c + \ldots,
\label{nAbGT} 
\end{eqnarray}
showing only the leading term of an infinite series in powers of $gf$, $f_{abc}$ being a structure 
constant of the gauge group.

For $SU(N)$ theories with $N \geq 2$ where $f_{abc} \neq 0$, the presence of non-Abelian terms dependent 
on $gf$ causes serious complications: First, just the first non-Abelian gauge term shown in 
Eq. (\ref{nAbGT}) depends on both $\Omega$ and $A$, allowing it to ``twist'' the internal structure of $A$ 
itself in different ways depending on the exact circumstances in every gauge transformation. Lorc\'e's 
revision misses the real culprit that is this troublesome non-Abelian gauge term, and wrongly blames the 
external frame rotation that is working properly. 

Second, non-Abelian vector potentials satisfy {\it nonlinear differential equations} (nLDE) that cannot 
accommodate the non-Abelian gauge transformation (\ref{nAbGT}) easily or even allow an easy solution of 
a chosen gauge definition. We shall return to describe this basic nonlinear obstacle more fully after 
first setting the stage by discussing the stated second objective of this paper.

\section{Dynamical and non-dynamical parts of the vector potential}
\label{sec: Dyn/Ndy}

The {\it linear differential equation} (LDE) (\ref{inhMaxEq}) also plays a central role in determining 
the origin of the {\it dynamical} (dyn) part of $A$ in ED: 
\begin{eqnarray}
A &=& A_{\rm dyn} + A_{\rm ndy}:   \nonumber \\
{\cal L} A_{\rm dyn} &=& j, \quad {\cal L}  A_{\rm ndy} = 0.
\label{dyn/nd}
\end{eqnarray}
That is, $A_{\rm dyn}$ is a particular solution of the inhomogeneous LDE with a non-zero 4-current 
$j$ that is gauge-independent. (Note that in QED, $j \equiv \overline{\psi}(x) \gamma\psi(x)$ is a local 
density where any arbitrary phases in the fermion fields always cancel in pair. The $\gamma^\mu$s are 
hL-invariant numerical 4 $\times$ 4 matrices acting on 4-component, hL-{\it variant} Dirac spinors 
$\psi(x)$. The spacetime ($\mu$ index) structure of the expression is designed to guarantee the covariant 
4-vector property of $j$ under hL transformation by the hL covariance of the Dirac equation, or vice versa 
\cite{Sakurai67}.) An additional {\it non-dynamical} (ndy) homogeneous solution $A_{\rm ndy}$ (for the 
equation with $j=0$ that is also gauge-independent) can be added to $A_{\rm dyn}$ to change the 4D 
boundary condition satisfied by their sum to some desirable value without changing the dynamics induced 
by the source 4-current $j$ that is already contained in a particular $A_{\rm dyn}$. This separation into
dynamical and non-dynamical parts is a {\it gauge-independent} process; it can and should be made before 
a choice of gauge.

Note that $A_{\rm dyn}$ contains the same dynamics as the original gauge-invariant but higher-ranked 
field tensor $F$. The non-dynamical part $A_{\rm ndy} = A - A_{\rm dyn}$ that is left must include the 
gauge term $\partial^\mu\Omega(x)$ of Eq. (\ref{GT}) because with $F^{\mu\nu} = 0$ the gauge term 
satisfies the homogeneous LDE. However, this LDE also has other solutions with nonzero $F^{\mu\nu}$ that
contain real physics, as we shall now discuss.

The dynamical/non-dynamical treatment does exact a price: A gauge degree of freedom now 
appears explicitly in the inhomogeneous LDE (\ref{inhMaxEq}), so that a choice of gauge is now required 
before Eq. (\ref{inhMaxEq}) can be solved in practice. The LDE itself takes different forms in different 
gauges, thus showing explicitly the variety of vector potentials that can be used to describe the same 
physics. In Lorenz gauges, both $A_{\rm ndy, L}$ and the residual gauge term $R_{\rm L}$ satisfy the same 
homogeneous wave equation. So homogeneous solutions of the gauge type exist in the Lorenz gauge. There 
are additional {\it physical} solutions with nonzero $F^{\mu\nu}$ in any gauge: In Lorenz gauges, they 
describe free electromagnetic waves in all spatial directions in 3D 
space. In the Coulomb gauge, the transversality condition excludes all pure-gauge solutions because they 
are longitudinal. One is then left with only ``physical'' solutions that satisfy wave equations in the 2D 
transverse momentum space and a time-like scalar potential $A^0$ that satisfies a Laplace equation \cite
{Jackson99}. 

There are infinitely many solutions for both types of $A_{\rm ndy}$ because of the infinite variety of
gauge functions $\Omega$ that appear in gauge solutions, and of boundary conditions for physical 
solutions. Since a dynamical solution remains a dynamical solution after the addition of any homogeneous 
solution, there are infinitely many $A_{\rm dyn}$ too. Furthermore, the multivalued $A_{\rm ndy, g}$ for
any gauge $g$ is already contained in the multivalued $A_{\rm dyn, g}$ and is not independent of it: 
If we enumerate the different values in these multivalued functions as $A_{\rm dyn, g}(x, n)$ and 
$A_{\rm ndy, g}(x, m, n)$ by adding counting numbers $m, n$ as additional arguments, then 
\begin{eqnarray}
A_{\rm ndy, g}(x, m, n) = A_{\rm dyn, g}(x, m) -  A_{\rm dyn, g}(x, n). 
\label{nd}
\end{eqnarray}
There is thus no additional final gauge transformation, gauge rotation or even boundary conditions of the 
types described in \cite{Lorce13, Leader14} for the simple reason that all solutions can be included in 
the multivalued object $A_{\rm dyn, g}$, previously called $A_{\rm g}$. Its different values have the same 
symmetry properties, especially that under hL transformations, differing only in numerical values.

The situation is similar to that in a much simpler problem: The 1D Newton equation $\ddot x(t) = j(t)$ 
has the multivalued solution $x(t) \equiv x_{\rm dyn}(t)$ containing the entire 2D infinity 
of numerical solutions for all possible choices of initial conditions (IC) from points of the 
2D IC space. A 2D infinity of force-free homogeneous solutions denoted collectively as $x_{\rm ndy}(t)$ 
also exist, but they can all be extracted from $x_{\rm dyn}(t)$ using a 1D version of Eq. (\ref{nd}). 
The gauge terms are absent, just like the Maxwell $A_{\rm C}$ in the Coulomb gauge. 

Of course, none of the other 2D infinity of multiple solutions is needed if the particular solution on 
hand is already the ``physical'' solution satisfying the desired ICs. So the word ``multivalued'' is 
used in this paper in a generic or familial sense and not in the literal sense that a physical state 
shows multiple realities. On the contrary, each completely specified physical state is described by only 
one of these multiple solutions. 

The multiple solutions of the vector potential $A$ play a more complex role, as we have already discussed. 
For example, the free electromagnetic waves contained in the physical part of the non-dynamical vector
potential $A_{\rm ndy}$ quantize to the infinite variety of free photon states with nonzero number of 
photons of different energies propagating in all directions in space from spatial infinity to spatial 
infinity. The multiple solutions of the dynamical vector potential $A_{\rm dyn}$ contain the additional 
dynamics associated with distinct choices of the 4-current $j$. For problems involving only the unique 
gauge field bound in a specific atomic state, an appropriate ``particular'' $A_{\rm dyn}$ solution can be 
defined without using any $A_{\rm ndy}$ term of the physical type, but the gauge part of $A_{\rm ndy}$ 
can still be kept to display explicit gauge invariance \cite{Chen08}. More generally however, the 
dynamical/non-dynamical treatment provides a more complete description of the physical contents of $A$.   

The gauge-invariant dynamical/non-dynamical separation of $A$ is also hL covariant, since it depends on 
the presence or absence of a covariant 4-vector current $j$. Eq. (\ref{inhMaxEq}) then retains its 
standard hL covariant tensor structure for all hL covariant 4-vector $A$, as intended in the original
covariant formalism: This result holds both for the gauge-preserving covariant 4-vector $A_{\rm cg}$ in 
a covariant gauge cg and for a gauge non-preserving but covariant 4-vector $A_{\rm vg}$ in the 
variable-gauge (vg) treatment of any other gauge choice. For the gauge-restored but non-covariant (nc) 
4-vector construct $A_{\rm nc}$ for a non-covariant gauge however, the additional gauge-restoring term 
$R'_{\Lambda C \rightarrow {\rm nc}}(x')$ destroys the explicit hL invariance of interaction Lagrangian. 
This defect does not prevent QED quantization in the Coulomb gauge in a {\it single frame} 
\cite{Bjorken65,Weinberg96}. 

So the answer to the objection \cite{Ji10} that a non-covariant gauge like the Coulomb gauge is not hL 
covariant is that in a gauge-independent and hL-invariant theory, the Coulomb gauge used in frame $x$ 
alone gives correct results because of either a hidden hL covariance or an actual hL covariance when used 
in a variable-gauge context. It is not possible to preserve {\it explicit} hL covariance when the gauge is 
{\it fixed} at the Coulomb gauge in all frames.  

Of course, covariant gauges in general, and Lorenz gauges in particular, are {\it special} because they 
preserve both their gauge and the full hL covariant structure of all expressions in all hL frames.
Nevertheless, it is the Coulomb (radiation or Landau) gauge that provides the simplest and physically 
most intuitive description of electromagnetic radiation or photons. The 3D space rotation invariance in 
its gauge condition allows the radiation/photon to travel in the same way along any spatial direction 
${\bf e}_{\bf k}$, while the explicit spacetime asymmetry in its gauge condition is designed to confine 
${\bf A}_{\rm C}$ entirely to the 2D subspace of transverse polarizations perpendicular to 
${\bf e}_{\bf k}$. An electromagnetic wave traveling in a definite direction ${\bf e}_{\bf k}$ described 
so nicely by the Coulomb gauge is an example of a commonplace phenomenon that the spacetime symmetry of a 
physical state can differ from the hL invariance of the underlying Lagrangian. Finally, the exclusion of 
the entire gauge degree of freedom actually means that ${\bf A}_{\rm C} = {\bf A}_\perp$ is gauge 
independent or invariant (\cite{CT89}, for example). 

For the inhomogeneous nonlinear differential equations (nLDE) satisfied by non-Abelian vector potentials, 
the separation of $A$ into parts
faces a serious obstacle: There are dynamical solutions $A_{{\rm dyn}, a}$ for a non-zero Dirac 4-current 
$j_a = \overline{\psi}\gamma t_a\psi$ in the inhomogeneous nLDE, and non-trivial non-dynamical solutions 
$A_{{\rm ndy}, a}$ for the associated homogeneous nLDE with $j_a = 0$. However, linear superpositions of 
$A_{{\rm dyn}, a}$ and $A_{{\rm ndy}, a}$, or decomposed parts of both do not in general satisfy nLDEs 
simply related to the original equations. 

One can use systematic perturbation theory \cite{Peskin95} however when there are no nonlinear instabilities 
or complications. An iterative perturbation theory can be set up by first writing all expressions with the 
dimensionless non-Abelian coupling constant $g$ shown explicitly in formulas. Then the non-Abelian nLDE for 
$A$ can be re-arranged so that all terms dependent on $g$ are moved to the RHS:
\begin{eqnarray}
\partial_\mu(\partial^\mu A_a^\nu - \partial^\nu A_a^\mu) 
&=&  -g(j_a^\nu + j_{2, a}^\nu) - g^2 j_{3, a}^\nu , 
\label{inhYMeq}
\end{eqnarray}
where the non-Abelian nonlinear terms $j_{2, a}, j_{3, a}$ containing 2 and 3 vector potentials respectively may
be treated as gauge 4-currents in an iterative solution: The calculation starts from a chosen unperturbed 
$A_{\rm ndy}$ (with $g = 0$ on the RHS). The calculated terms to order $n$ can then be used in the 
nonlinear 4-currents on the RHS to drive the solution to order $n+1$ in the resulting linearized DE. In 
this linearized theory, Abelian quantizations in the usual gauges used in Abelian theories and Feynman 
diagrams can be used. This perturbative method is different from the procedure suggested in \cite{Chen08}. 

Such perturbation methods may not always work well because physical states may contain very significant 
components where one or more linearized gauge bosons appear when $g$ is large. Certain collective 
phenomena may require a great deal of effort to describe.  

The nonlinearity of non-Abelian gauge theories causes further complications \cite{Itzykson80, Coleman85, 
Peskin95, Rajaraman82}. The non-Abelian fermion 4-current $j_a \equiv \overline{\psi}(x) \gamma t_a\psi(x)$ 
and field tensor $F_a$ are both {\it gauge-dependent}, and more difficult to handle. Unlike the 
simple mathematical structure allowed by the linear superposition property of Abelian theories, 
non-Abelian nonlinearity admits a multitude of structures at the classical level: A gauge condition may 
have multiple solutions called Gribov copies, or no solution at all. The non-Abelian nLDE (\ref{inhYMeq}) 
with 4-current $j=0$ satisfied by $A_{\rm ndy}(x)$ of pure gauge can be used to define a nonlinear 
(with $g \neq 0$) classical vacuum that has infinitely many distinct solutions of different internal 
structures characterized by different topological winding numbers or kinks $-\infty \leq n \leq \infty$. 
This topological structure for all simple Lie groups, including $SU(N \geq 2)$, turns out to be the same 
as that for their $SU(2)$ subgroup alone, according to Bott's theorem. 

The perturbation theory sketched in Eq. (\ref{inhYMeq}) can be used to start from any nonlinear 
solution $A_{{\rm ndy}, n}(x)$ to give an approximate iterative-perturbative solution with the same 
winding number $n$ (or in the same homotopy class) as $A_{{\rm ndy}, n}(x)$. The resulting linearized 
perturbation theory can accomodate second quantization into gauge bosons based on the topological vacuum 
state $|n\rangle$ centered around the classical $n$-vacuum. Quantum tunnelings between neighboring 
$|n\rangle$-vacua give rise to transient events lasting only instances in time called instantons and 
anti-instantons, instantons' time-reversed twins. 

Quantization confers the non-native ability of linear superposition: The true vacuum that includes 
quantum couplings between $|n\rangle$-vacua is one of the $\theta$-vacua 
$|\theta\rangle = \sum_n e^{in\theta}|n\rangle$, where the arbitrary Bloch phase $n\theta$ appearing 
with a quantum $|n \neq 0\rangle$-vacuum is $n$-dependent, as required by the periodic appearance of the 
degenerate $|n\rangle$-vacua in the 1D winding number space. For $\theta$-independent Lagrangians and for 
certain gauges, these $\theta$ vacua can be considered disjoint and duplicate mathematical realizations of 
the same physical vacuum.

The native nonlinearity persists even among the quantized gauge bosons however, for they clump together 
with or without interacting fermions into clusters of zero total non-Abelian charge $g$. At ultra-short 
distances, the particles inside each cluster have been found unexpectedly to be free and non-interacting, 
thus leading to the unfamiliar situation that the physical picture gets progressively simpler as the 
distance scale of observation decreases. Finally, gluons of nonzero charge $g$ cannot propagate freely in 
free space. Hence some of the non-dynamical but physical solutions of $A_{\rm ndy}$ of the current-free 
nonlinear Yang-Mills equations for $A$ giving nonzero $F^{\mu\nu}$ quantize to physical states of 
glueballs of total $g=0$ propagating freely in free space.

I thank Fan Wang and Cedric Lorc\'e for many helpful discussions.

\end{document}